\documentclass[11pt]{article}
\usepackage[normalem]{ulem}
\pdfoutput=1 % if your are submitting a pdflatex (i.e. if you have

\usepackage{jheppub} % for details on the use of the package, please
\usepackage{url}
\usepackage{caption}
\usepackage{subcaption}
\usepackage{extarrows}
\usepackage{multirow}

\usepackage[latin9]{inputenc}
\usepackage{float}
\usepackage{amsmath}
\usepackage{amssymb}
\usepackage{graphicx}
\usepackage{esint}
\usepackage{hyperref}
\usepackage{comment}
\usepackage{color}
\usepackage{microtype}
\usepackage{cleveref}
\usepackage{breakurl}
\usepackage{bbm}
\newcommand{\be}{\begin{equation}}
\newcommand{\ee}{\end{equation}}
\newcommand{\ben}{\begin{displaymath}}
\newcommand{\een}{\end{displaymath}}
\newcommand{\bea}{\begin{eqnarray}}
\newcommand{\eea}{\end{eqnarray}}
\def\K{K{\"a}hler }
   \newcommand{\rf}[1]{(\ref{#1})}
\newcommand{\vp}{\varphi}

\def\be{\begin{equation}}
\def\ee{\end{equation}}
\def\bea{\begin{eqnarray}}
\def\eea{\end{eqnarray}}
\def\ba{\begin{array}}
\def\ea{\end{array}}
\def\bit{\begin{itemize}}
\def\eit{\end{itemize}}

\def\a{\alpha}

\def\vp{\varphi}

 \makeatletter

\allowdisplaybreaks

\makeatother

\makeatletter
\DeclareRobustCommand{\rcite}[1]{%
  \rcite@aux#1,\@nil{#1}%
}
\def\rcite@aux#1,#2\@nil#3{%
  \if\relax#2\relax
    Ref.~\cite{#3}%
  \else
    Refs.~\cite{#3}%
  \fi
}
\makeatother

\hypersetup{
    colorlinks = true,
    citecolor = {blue},
    linkcolor = {blue},
    urlcolor = {blue},
}

 \title{\rm { \huge  \bf   Singular   {\boldmath $\a$}-attractors   }}

\author{Renata Kallosh \ and }
\author{Andrei Linde}

\affiliation{Leinweber Institute for Theoretical Physics and Department of Physics,\\ Stanford University, Stanford, CA 94305, USA}
\emailAdd{kallosh@stanford.edu}
\emailAdd{alinde@stanford.edu}

\abstract{Inflationary $\alpha$-attractor models naturally appear in supergravity with hyperbolic geometry. The simplest versions of  $\alpha$-attractors, {\it T- and E-models,} originate from theories with non-singular potentials.  In canonical variables, these potentials have a plateau that is approached exponentially fast at large values of the inflaton field $\varphi$. In a closely related class of polynomial $\alpha$-attractors, or {\it P-models}, the potential is not singular, but its derivative is singular at the boundary. The resulting inflaton potential also has a plateau, but it is approached polynomially. In this paper, we will consider a more general class of potentials, which can be singular at the boundary of the moduli space, {\it S-models}.  These potentials may have a short plateau, after which the potential may grow polynomially or exponentially at large values of the inflaton field. We will show that this class of models may provide a simple solution to the initial conditions problem for $\alpha$-attractors and may account for a very broad range of possible values of $n_{s}$ matching the recent ACT, SPT, and DESI data.}

\begin{document}

\maketitle

 \parskip 7pt

% \tableofcontents{}
%\newpage

\section{Introduction}

More than a decade ago, observational data from WMAP and Planck \cite{Hinshaw:2012aka,Planck:2013jfk} attracted attention to two different inflationary models that matched the data particularly well: the Starobinsky model \cite{Starobinsky:1980te} and the Higgs inflation \cite{Salopek:1988qh,Bezrukov:2007ep}. These models differ significantly from each other, yet both predict the same values for the spectral index $n_{s}$ and tensor-to-scalar ratio $r$ as functions of the number of  e-foldings $N$ in approximation when $N$ is large.

Later, a theory of cosmological attractors was discovered, such as conformal attractors \cite{Kallosh:2013hoa} and $\xi$-attractors \cite{Kallosh:2013tua}. These  have the same predictions for $n_{s}$ and $r$ as Starobinsky and  Higgs inflation, across a broad class of inflationary potentials. The next step was made with the invention of $\alpha$-attractors \cite{Ferrara:2013rsa,Kallosh:2013yoa}, which made the same prediction for $n_{s}$, but allowed to dial any desirable value of $r$ by a choice of the parameter $\alpha$ related to the hyperbolic geometry of the moduli space. We called these models ``attractors'' because under certain conditions their predictions are stable with respect to significant modifications of the inflationary potential.

 All of these models provide a good fit to currently available CMB data, including the combination of Planck and the latest ACT \cite{AtacamaCosmologyTelescope:2025blo} and SPT data, giving $n_{s} = 0.9684 \pm 0.0030$ \cite{SPT-3G:2025bzu}, though even in this case there is a large dispersion of the constraints on $n_{s}$ depending on the choice of the latest Planck maps-likelihood combination\cite{Jense:2025wyg}.
But once the recent DESI results \cite{DESI:2025zgx} are taken into account \cite{AtacamaCosmologyTelescope:2025blo,SPT-3G:2025bzu}, the spectral index for CMB-SPA + DESI becomes higher by about  2$\sigma$: $n_{s}=0.9728\pm0.0027$, according to  \cite{SPT-3G:2025bzu}. 
It is not easy to make the benchmark inflationary models discussed above compatible with these results. 

As emphasized in \cite{SPT-3G:2025bzu,Ferreira:2025lrd}, one should be very careful when interpreting the results of combining CMB and DESI data, as these datasets are at about $2\sigma$ - $4\sigma$ tension with each other.  Moreover, the problem disappears altogether in the two-field $\alpha$-attractor models, such as hybrid attractors. In these models, one can have much higher values of $n_{s}$  because of the uplift of the $\alpha$-attractor potential of the inflaton by the second field \cite{Kallosh:2022ggf, Braglia:2022phb}. In this paper, we will consider other modifications of the original single-field $\alpha$-attractors.

The main reason for the universality and stability of the predictions of the original versions of $\a$-attractors, {\it T- and E-models}, is the assumption that the scalar field potential and its derivatives are non-singular at the boundary of the moduli space \cite{Kallosh:2013yoa,Kallosh:2016gqp,Kallosh:2022feu}. This condition implies that the potential of the canonically normalized inflaton field $\vp$ exponentially fast approaches a plateau. One may call such models exponential attractors. 

On the other hand, if one relaxes this assumption just a little and considers the potentials that are regular at the boundary but have singular derivatives, the attractor regime changes, and the potential approaches the plateau according to a power law \cite{Kallosh:2022feu}. These $\alpha$-attractors are called polynomial, or  {\it P-models}. The values of $n_{s}$ in P-models can be significantly higher than in the original versions of $\a$-attractors, matching the CMB-DESI bound $n_{s}=0.9728\pm0.0027$ \cite{Kallosh:2022feu,Kallosh:2025ijd}.

As a next step, one may consider the possibility that the potentials are singular at the boundary of the moduli space. We will call such models {\it S-models}.  The singular terms in the potential lead to a power-law or exponential uplift of the plateau at very large values of the inflaton field.  S-models have been discussed in the past in \cite{Linde:2017pwt,Linde:2018hmx}, where it was found that this uplift may significantly simplify the solution of the problem of initial conditions in the $\alpha$-attractor models. In addition, as we will see shortly, in this class of potentials one may increase $n_{s}$ to any desired value.

In this paper, we will describe S-models \cite{Linde:2017pwt,Linde:2018hmx} and provide an interpretation of several other recently discussed inflationary models in terms of $S$-model versions of $\alpha$-attractors where singular terms of the form $\delta\, V_{\rm sing}$ are added to the potential while the hyperbolic geometry of the kinetic terms is preserved. The way the new terms $ V_{\rm sing}$ affect the attractors can be seen in the new properties of $n_s$:  the deviation of $n_s$ from its universal  $\a$-independent  value $n_s =1-2/N$ is proportional to the coefficient $\delta$ in front of the term $V_{\rm sing}$. Since the deviation required to match the recent CMB-DESI data is relatively small \cite{AtacamaCosmologyTelescope:2025blo,SPT-3G:2025bzu}, it can be achieved by introducing the terms  $\delta\, V_{\rm sing}$ with a very small coefficient $\delta$.

\section{Common properties of T- , E- , P- and S-models}

A common feature of all $\a$-attractors embedded in supergravity is that these models have {\it kinetic terms associated with hyperbolic  geometry} \cite{Kallosh:2013yoa,Cecotti:2014ipa,Kallosh:2015zsa,Carrasco:2015uma}.

In disk variables $Z$  or in half-plane variables $T$  the \K potentials are 
\be
 K(Z, \bar Z)=-3\a \log(1-Z\bar Z), \,  \qquad K(T, \bar T)=-3\a\log(T+\bar T) \ .
\label{K}\ee
One can represent $Z$ and $T$ as
\be
Z = \tanh {\vp\over \sqrt {6\a}}  e^{i\, \theta}, \, \qquad  T =  T= e^{-\sqrt{2\over 3\a}\vp}+i \, \theta \ ,
\ee
where $\vp$ is a canonically normalized inflaton field and $\theta$ is the axion.
The Cayley relation between disk and  half-plane variables of hyperbolic geometry is
\be
T= {1+Z\over 1-Z}\, ,  \qquad Z= {T-1 \over T+1} \ .
\label{cayley}\ee
The corresponding  kinetic terms are
\be
K_{T \bar T} \partial T \partial \bar T= {3\alpha\over 4} \, {\partial T \partial \bar T\over ({\rm Re} \,  T )^2}= {3\alpha} \, {\partial Z \partial \bar Z\over (1-Z\bar Z)^2} = K_{Z\bar Z} \partial Z \partial \bar Z \ .
\label{kin}\ee
  This  
 metric corresponds to a symmetric space ${SU(1,1)\over U(1)}$ with constant negative \K curvature  $R_K= - {2\over 3\a}$. Thus, parameter $\a$ defines 
the curvature of the \K manifold.
 T-models are based on disk variables $(Z, \bar Z)$, E-models are based on half-plane variables $(T, \bar T)$. 
The boundary of the moduli space in disk variables 
$(Z, \bar Z)$ and in half-plane variables $(T, \bar T)$ is defined as follows
\be
Z\bar Z \to 1\, ,  \qquad T+\bar T \to 0 \ .
\label{boundary}\ee
The potentials of exponential and polynomial $\a$-attractors are regular at the boundary of the moduli space. In the exponential case, the first derivative of the potential is regular, whereas in polynomial $\a$-attractors it is singular. Polynomial attractors have higher values of $n_s$, which make them suitable for describing the recent CMB-DESI data.

The new S-models of $\a$-attractors we propose here are based on potentials which have some small terms $\delta  \, V_{\rm sing}$  which are singular at the boundary of the moduli space. These are absent in the plateau models, exponential and polynomial $\a$-attractors. These terms uplift the plateau at large values of the inflaton field, but if $\delta$ is sufficiently small, the shape of the potential at smaller values of the inflaton field is preserved.  
Since to fit the new data we need to deviate from universal attractor values by only 2 or 3 $\sigma$, one can achieve the desirable result by adding the new terms with a very small coefficient $\delta \ll 1$.

To summarize, the main difference between T- , E- , P-, and S-models is in the properties of their potentials at the boundary of the moduli space defined in eq. \rf{boundary}.

T-  and  E-models have potentials that are regular at the boundary.

P-models have non-singular potentials with derivatives that are singular at the boundary.

S-models have potentials that are singular at the boundary.

\noindent Complete supergravity versions of $\a$-attractors with an arbitrary potential are presented in \cite{Kallosh:2025dac} and in Appendix A of this paper.

\section{T- , E- and P-models with   plateau potentials}
In the simplest versions of  $\alpha$-attractors ({\it T-models and E-models}), the potentials and their derivatives are required to be regular functions of their $(Z, \bar Z)$ or $(T, \bar T)$ variables at the boundary of the moduli space defined in eq. \rf{boundary}. The simplest T-model and E-model potentials are 
\be\label{sm}
V_{T}=V_0 (Z\bar Z)^n  \ , \qquad V_{E}=V_0 \, \Bigl(1- {T+\bar T\over 2}\Bigr)^{2n}   \ .
\ee
These potentials  have a plateau behavior near the boundary at $Z\bar Z \to 1$
and $T+\bar T \to 0$.
Note that these simple potentials do not depend on the axion field $\theta$. To simplify the cosmological models, one may want to stabilize the axions $\theta$ at $\theta = 0$, so that  $Z=\bar Z$ and $T=\bar T$. This is not always necessary \cite{Achucarro:2017ing, Linde:2018hmx}, but it is always possible to do so by various methods, without affecting the potential in the inflaton direction, see e.g.  \cite{Kallosh:2010xz,Kallosh:2017wnt,Kallosh:2021fvz,Carrasco:2025rud,Kallosh:2025jsb}. In particular, in the recently developed streamlined supergavity context \cite{Kallosh:2025dac} one can construct cosmological models with {\it any} \K potential and {\it any} desired potential, including potentials with a stabilized axion field $\theta = 0$, without affecting the potential of the inflaton field $\vp$, see Appendix A of this paper. 

The potentials of the   simple T-models and E-models  \rf{sm} with respect to the canonically normalized inflaton field $\vp$ are
\be
V_{T}=V_0\, \tanh^{2n} {\vp\over \sqrt{ 6 \a}} \, , \qquad V_{E}=V_0 \left (1- e^{-\sqrt{2\over 3\a}\vp}\right )^{2n}  \ .
\ee
Both potentials exhibit plateau behavior at $\vp \to +\infty$, but the T-model potential, being symmetric under $\vp \to -\vp$, also has a plateau at $\vp \to -\infty$. 
 Both models have an exponential approach to a plateau at large positive canonical fields $\vp$
\be
V_{\rm exp}(\vp)|_{\vp \to + \infty}= V_0( 1-e^{-{\sqrt{2\over 3\a} \vp}}+\cdots)\, ,
\label{exp}\ee
We call these theories exponential $\a$-attractors. They have simple predictions at the attractor point  in terms of the number of e-foldings $N$  for large $N$   \cite{Kallosh:2013yoa}
\be
n_{s} = 1-{2\over N} \ , \qquad  r = {12 \a\over  N^{2}} \ .
\label{a}\ee
 Moreover, one may also consider general potentials $V(Z, \bar Z)$ with a stabilized axion that have a minimum at $Z = 0$.  We will assume that these potentials grow with an increase of $|Z|$, and are finite and have finite derivatives at the boundary of the moduli space $Z\bar Z = 1$. Then one can show that all such theories have the same predictions in the large-$N$ limit. That is why we called these theories ``attractors.''

{\it P-models}  have the same \K geometry and kinetic terms as T- and E-models in eqs. \rf{K} and \rf{kin}. Potentials of the P-models are continuous at the boundary of the moduli space   $Z\bar Z \to 1$
or $T+\bar T \to 0$, but the derivatives of the potentials diverge at the boundary. 

Consider the P-model potential in half-plane variables of the form
\be
V_{P}=V_0 {1\over 1+ \big(\ln {T+\bar T\over 2}\big)^{-2n}  } \ .
\ee
Near the boundary,  the potential is non-singular and has a plateau.
 However, the derivatives of this potential with respect to $T$ or $\bar T$ at the boundary $T+\bar T\to 0$ are divergent. When we switch to canonical variables at $T=\bar T$ where 
$T= e^{-{\sqrt{2\over 3\a}} \vp}$  we find the potential of the P-model is a KKLTI potential \cite{Martin:2013tda} with $\mu^2= {3\a/ 2}$
\be
V_P=V_0 {1\over 1+ \Big ({\sqrt{2\over 3\a}} \vp\Big)^{-2n} }= V_0 {\vp^{2n} \over ({3\a\over 2})^n+\vp^{2n} } \ .
\ee
These models have a polynomial approach to a plateau at large positive canonical fields $\vp$:  
\be
V_P=V_0 \Bigl(1- \Big({\sqrt {3\a/ 2} \over \vp}\Big )^{2n}  +\cdots \Bigr) \, 
\ee
but near the minimum at $\vp=0$ the potentials behave as $\vp^{2n}$. A slight generalization of these potentials allows us to use any positive non-integral values of $n$ \cite{Kallosh:2022feu}.

P-models have the attractor predictions for the cosmological observables at fixed large $N$ where $k=2n$
\be\label{kalpha}\
  n_{s} = 1-{2(k+1)\over (k+2)N} \ ,   \qquad r = 8 \, k^2 \Big({3\a\over 2}\Big )^{k\over k+2} \Big({1\over  k  (k+2) N)}\Big)^{\frac{2 (k+1)}{k+2}} \ .
\ee  
By changing $k$ from 2 to $\infty$, one can describe a broad range of values of $n_{s}$ from $1-2/N$ to $1-3/2N$. For $N \sim 55$, this range is $0.967 < n_{s} < 0.973$. 
By including fractional $n$, so that $0<k<2$, one can extend the full range of $n_{s}$ to $1-2/N < n_{s }< 1-1/N$. For $N = 55$, the full range becomes $0.967 < n_{s} < 0.982$. This is more than enough to describe the recent CMB+DESI constraints $n_{s}=0.9728\pm0.0027$.

As we will see, in the S-model context, one can extend the range of possible values of $n_{s}$ even further, all the way to $n_{s}  = 1$.

\section{S-models}\label{singul}
\subsection{General case and known examples}\label{Sec:gen}
S-models have \K geometry in eq. \rf{K},  kinetic terms in eq. \rf{kin} and they have potentials with singularities at the 
boundary. 
 Examples of singular terms we add to standard attractor  potentials with a small coefficient $\delta$ are
\be
 {F(Z, \bar Z)\over (1-Z\bar Z)^k}\, , \qquad   G(Z, \bar Z) \ln^{k}  { (1-Z\bar Z)} \ .
\label{generalZ}\ee
 In half-plane variables, these can be 
\be
 {L(T, \bar T)\over (T+\bar T)^{k}}\, , \qquad  S(T, \bar T)  \ln^{k}  { (T+\bar T)} \ .
\label{generalt}\ee
Here, all functions $F(Z, \bar Z), G(Z, \bar Z), L(T, \bar T), S(T, \bar T)$ are regular at the boundary.
\begin{figure}[H]
\centering
		 \includegraphics[width=0.48\textwidth]{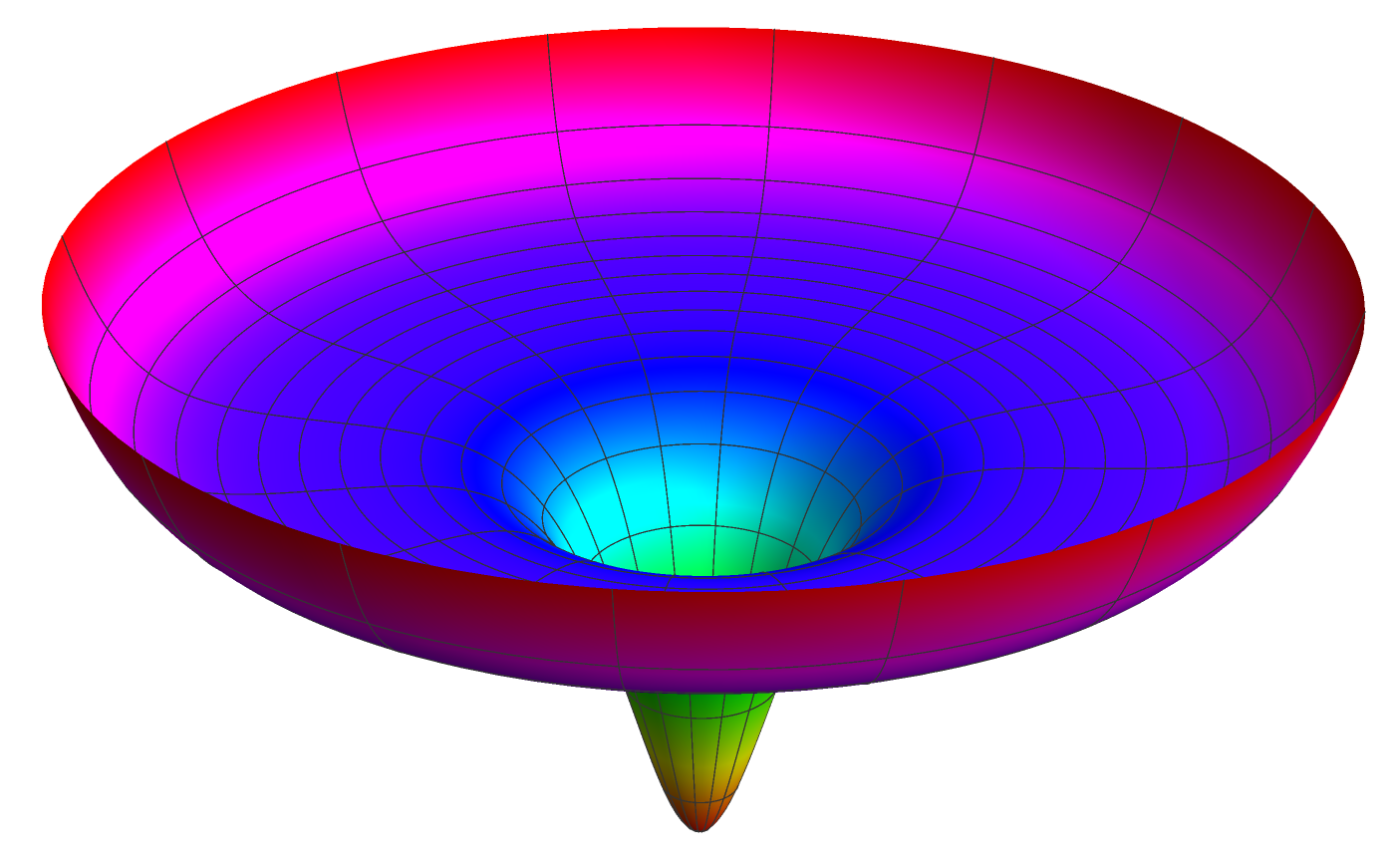}
        \caption{This figure from \cite{Linde:2018hmx} shows the axially symmetric  $\alpha$-attractor  plateau potential bounded by an exponentially steep wall, which emerges because of the singularity of the potential \rf{singL} at $|Z| = 1$. }
        \label{Wall}
\end{figure}
 Some of these singular S-models were already studied in 
\cite{Linde:2017pwt,Linde:2018hmx}, where  $\a$-attractors with a short plateau due to a singular term in the potential
were investigated. It was the case in \rf{generalZ} with $F(Z, \bar Z)=Z\bar Z  $ and $k=1$ so that the total potential is
\be
V(Z, \bar Z) = V_0\ Z\bar Z \, \Big ( 1   + \, {\delta \over (1-Z\bar Z)} \Big ) \ .
\label{singL}\ee
The potential does not depend on the axion field, so it is axially symmetric, but it does depend on the inflaton field $\vp$:
\be
V(\vp)=V_0 \left( \tanh^{2} \frac{\vp}{\sqrt{6\alpha}}  +  \delta \sinh^{2}   \frac{\vp}{\sqrt{6\alpha}} \right)~.
 \label{toymodelpotentialhyper}
 \ee
 For $\delta \ll 1$, this potential as a function of $Z=\tanh{\vp\over \sqrt{6\a}}\   e^{i\theta}$ has a long plateau, which ends at a very large $\vp$, where the potential
becomes exponentially steep, see Fig.  \ref{Wall}  \cite{Linde:2018hmx}.

\begin{figure}[H]
\centering
		 \includegraphics[width=0.56\textwidth]{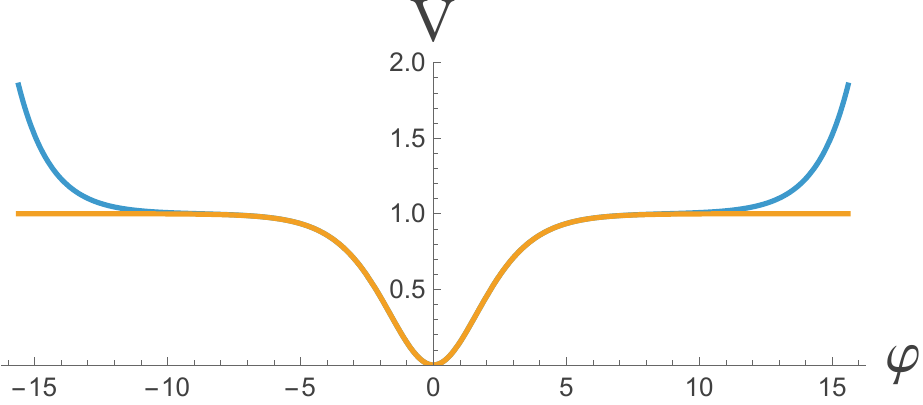}
        \caption{The blue line shows the potential \rf{toymodelpotentialhyper} for $\alpha = 1$, $\delta = 10^{{-5}}$ \cite{Linde:2017pwt}. The potential height is shown in units of $V_{0}$.  For comparison, the yellow line shows the usual $\alpha$-attractor with $\delta = 0$.}
        \label{finitePla}
\end{figure}

When the axion is stabilized, $Z=\bar Z =  \tanh{\vp\over \sqrt{6\a}}$, the inflaton potential becomes steep in the axion direction, but the potential in the $\vp$ direction remains the same as in equation \rf{toymodelpotentialhyper}. 
It is shown in Fig. \ref{finitePla}, see also \cite{Linde:2017pwt}.
In the next sections, we will describe several different models of this type, discuss the observational consequences of these models, and explain how the singular terms may help to solve the problem of initial conditions in this class of $\alpha$-attractors.\footnote{Note that in our models, the singular term appears with a very small coefficient $\delta$. Singular potentials with the singularity {\it of the entire potential} were previously introduced  
in \cite{Terada:2016nqg}. As noted in  \cite{Terada:2016nqg}, such potentials do not seem to offer advantages with respect to the interpretation of the observational data.  We are grateful to T.~Terada for the corresponding discussion.}

\section{Singular $\alpha$-attractors and observations}\label{obs}

As we already mentioned, some singular $\alpha$-attractor models have already been studied in \cite{Linde:2017pwt,Linde:2018hmx}. In this section, we will discuss these models, as well as their versions with a less singular (logarithmic) behavior.
\subsection{Logarithmic singularity}\label{ls}

\subsubsection{Singular T-models}\label{sst}
\begin{figure}[H]
\begin{center}
\includegraphics[scale=0.25]{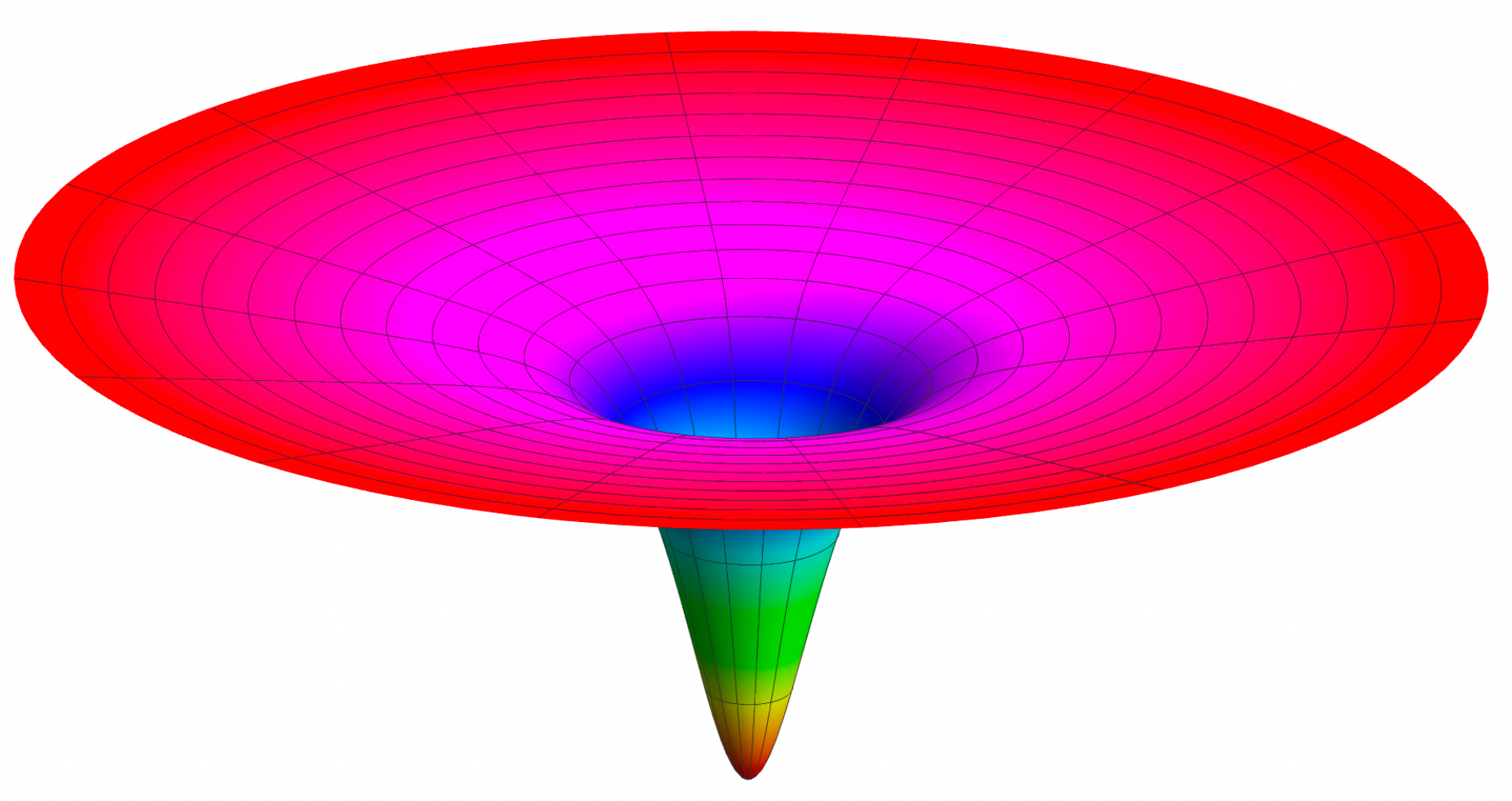}
\end{center}
%\vskip -0.5cm 
\caption{\footnotesize The potential \rf{sing1}  for $\alpha = 1/3$, $m,n=1$, and $\delta = 3\times 10^{-2}$}
\label{Fa}
\end{figure}

\begin{figure}[H]
\begin{center}
\includegraphics[scale=0.4]{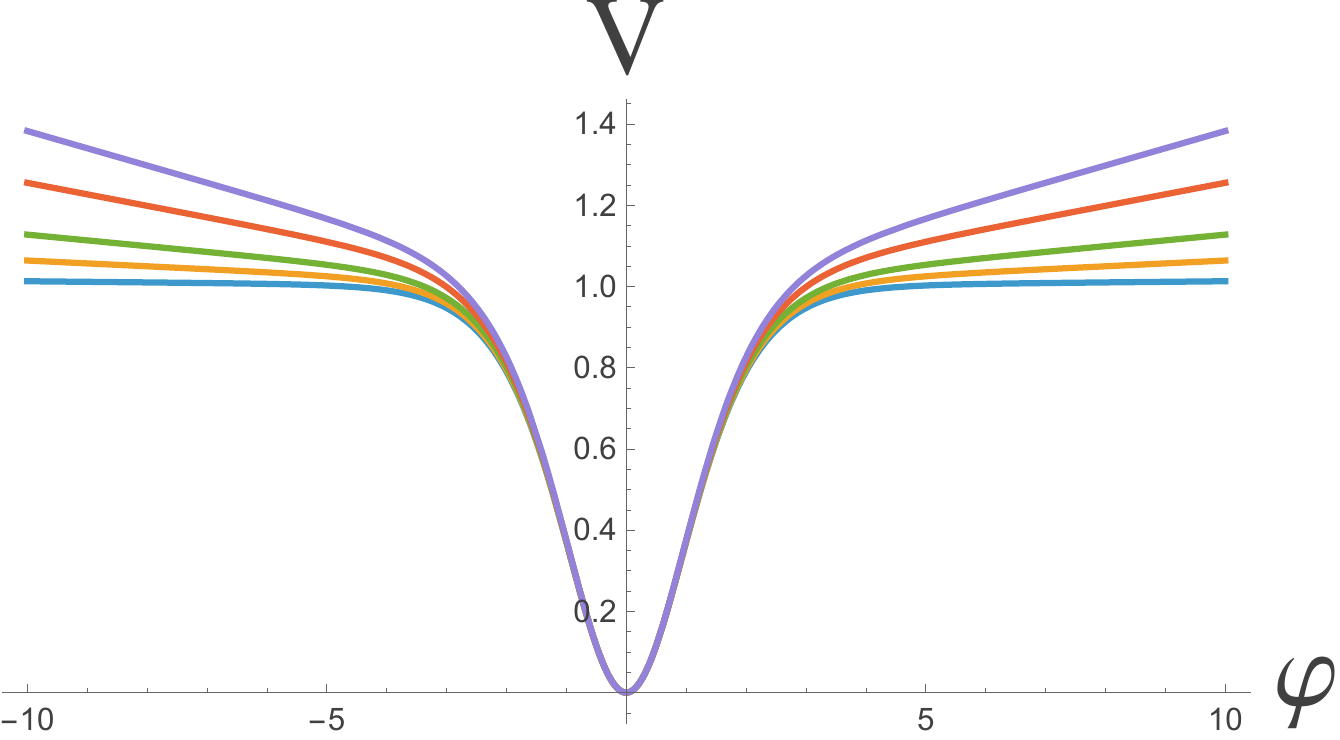}
\end{center}
%\vskip -0.5cm 
\caption{\footnotesize The potential \rf{sing2} shown in units of $V_{0}$  for $\alpha = 1/3$,  $m,n=1$, and $\delta = 3\times 10^{-2}$ (upper curve), $\delta=10^{-2}$, $\delta = 2\times 10^{-2}$,  $\delta=5\times 10^{-3}$ and $\delta= 10^{-3}$ (lower curve). }
\label{Fb}
\end{figure}

\be
V(Z, \bar Z) = V_0\ (Z\bar Z)^{m} \, \Big ( 1  + \, { \delta\  \ln^{n}(1-Z\bar Z)^{{-1}}} \Big ) \ .
\label{sing1}\ee
In canonical variables $\vp$, where $Z = e^{i \theta} \tanh {\vp\over \sqrt{6\alpha}}$, this simple potential reads
\be
V= V_0 \, \tanh^{2m} \big({\vp\over \sqrt {6\alpha}}\big) \Big ( 1  + \,   \delta\, \ln^{n}(\cosh^{2}({\vp/ \sqrt {6\alpha}}) \Big ) \ .
\label{sing2}\ee
This potential does not depend on $\theta$, as shown in Fig. \ref{Fa} for  $\alpha = 1/3$,  $m = n = 1$, $\delta = 3\times 10^{-2}$.
As we already mentioned, one can study inflation directly in the model \rf{sing1}, where the potential depends only on $\vp$ and not on the axion. Thanks to some properties of the hyperbolic geometry, all slow-roll inflationary motion in this class of models occurs in the radial direction, and axion perturbations do not contribute to adiabatic perturbations \cite{Achucarro:2017ing, Linde:2018hmx}.
Alternatively, one can modify the potential to stabilize the axion field $\theta$ at $\theta = 0$ without affecting the potential of the inflaton field $\vp$ \cite{Kallosh:2010xz,Achucarro:2017ing,Linde:2018hmx,Carrasco:2025rud,Kallosh:2025dac}.  In either case, one can ignore $\theta$, plot the potential \rf{sing2} as a function of the inflaton field $\vp$ (see Fig. \ref{Fb}), and calculate $n_{s}$ and $r$ for various $\delta$ (see Table \ref{k1}).  All numerical results for $n_{s}$ and $r$ in section \ref{obs} are given for $N = 55$. This potential has a simple power law behavior for $|\vp| \gg \sqrt{3\alpha/2}\  \delta^{-1/n}$. For any $m$ one has
\be
V= V_0\, \delta\, \Bigl({2\over 3\alpha}\Bigr)^{n/2}\, |\vp|^n \ .
\label{sing3}\ee
 For $m = 1, n = 2$ this potential at large $\vp$ becomes
\be
V= V_0\, \delta\, {2\over 3\alpha}\, \vp^{2} \ .
\label{sing5}\ee
In particular, for $n = 1$ and  $|\vp| \gg \sqrt{3\alpha/2}\, \delta^{-1}$
\be
V= V_0\, \delta\, \sqrt{2\over 3\alpha}\, |\vp| \ .
\label{sing4}\ee
\begin{table}[H]
\begin{center}
\begin{tabular}{ |p{1cm}||p{1.5cm}|p{1.5cm}|p{1.5cm}|p{1.5cm}|p{1.5cm}|  }
 \hline
 $\delta$ & $10^{{-3}}$ &$5\times 10^{-3}$ & $10^{-2}$ & $2\times 10^{-2}$ & $3\times 10^{-2}$\\
 \hline
 $n_{s}$ & 0.9651 & 0.9719 & 0.9784 & 0.9863& 0.9897\\
 \hline
 $r$ & 0.0015 &  0.0022 & 0.0035 & 0.0069 & 0.0114\\
 \hline
 \end{tabular}
\end{center}
\caption{\footnotesize Values of $n_{s}$ and $r$ for the model \rf{sing2} with $m,n=1$, $\alpha = 1/3$,   $N = 55$  and various $\delta$.}
\label{k1}
\end{table}
\begin{figure}[H]
\begin{center}
\includegraphics[scale=0.25]{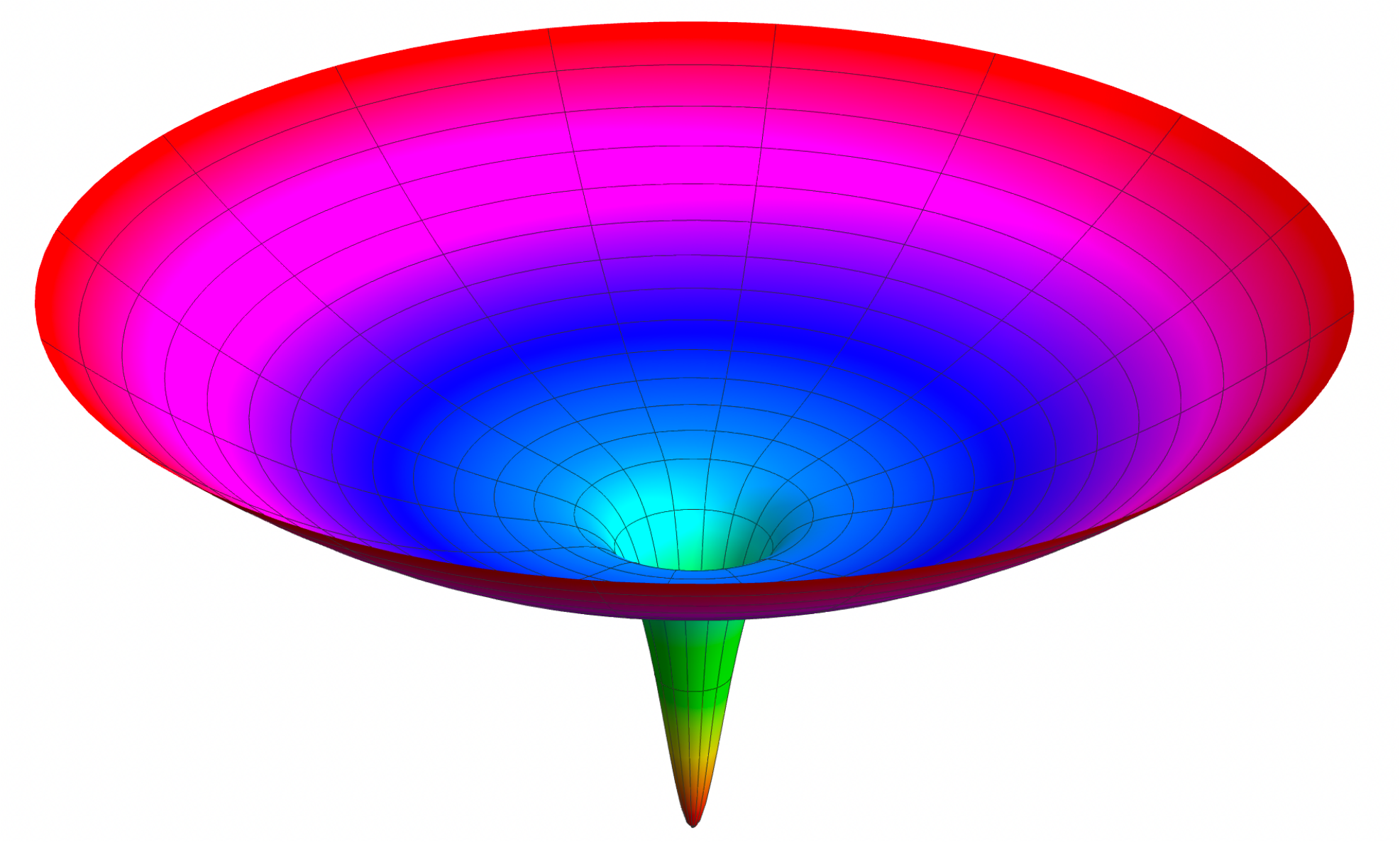}
\end{center}
%\vskip -0.5cm 
\caption{\footnotesize The potential \rf{sing2} shown in units of $V_{0}$  for $\alpha = 1/3$,  $m=1, n=2$, and $\delta =  10^{-3}$. To make its parabolic shape at large $\vp$ more visible we show the potential for $\vp < 20$. }
\label{Fchaotic}
\end{figure}
The shape of the potential \rf{sing1} for $\theta = 0$, $m = 1$, $n=2$ is shown in Fig. \ref{Fchaotic} for $\delta = 10^{{-3}}$ and $\vp < 20$.
The potential  \rf{sing1} for $m = 1$, $n=2$ after stabilization of the axion field $\theta = 0$ \rf{sing2}  is shown in Fig. \ref{Fchaotic} for $\delta = 10^{{-3}}$ and $\vp < 10$.
\begin{figure}[H]
\begin{center}
\includegraphics[scale=0.45]{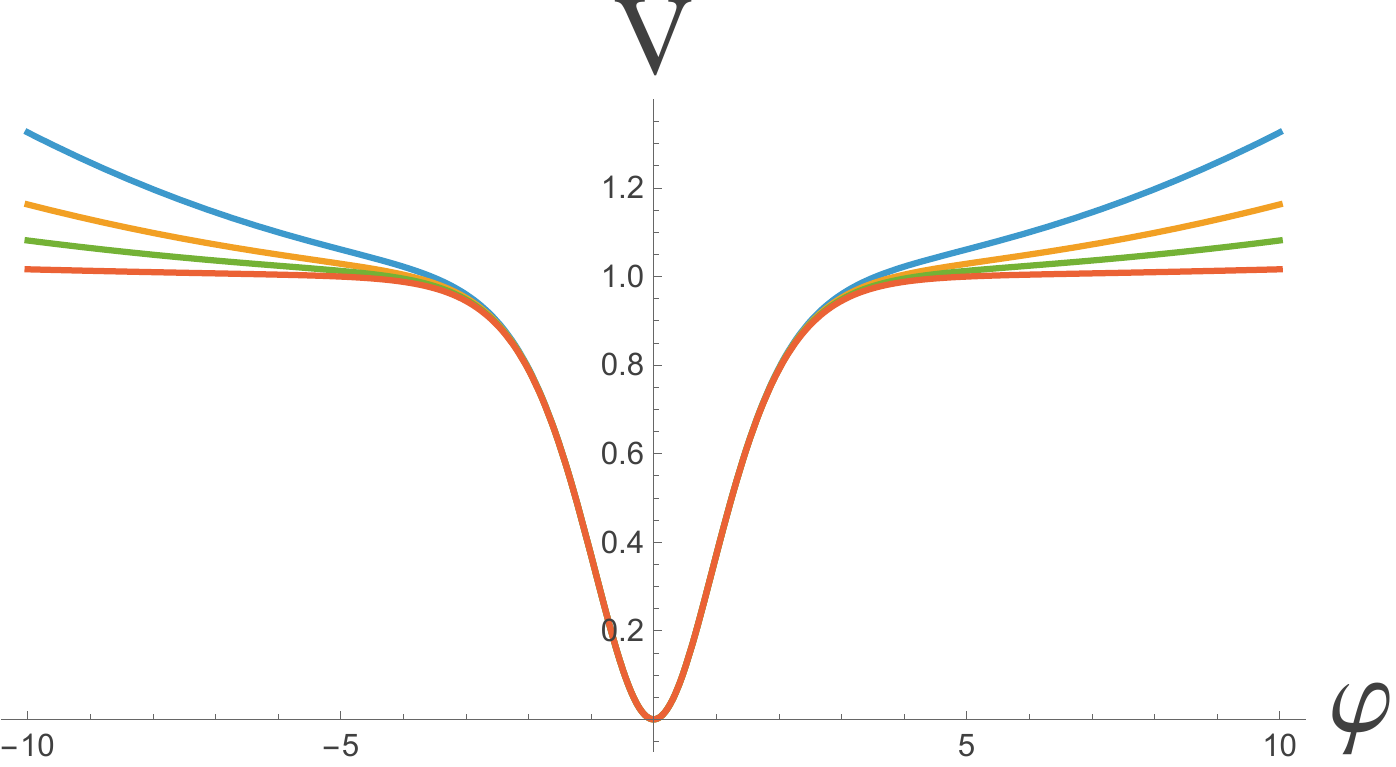}
\end{center}
%\vskip -0.5cm 
\caption{\footnotesize The potential \rf{sing2} is shown in units of $V_{0}$  for $\alpha = 1/3$,  $m=1, n=2$, and $\delta = 2\times 10^{-3}$ (upper curve), $\delta =10^{-3}$, $\delta=5\times 10^{-4}$ and $\delta= 10^{-4}$ (lower curve). }
\label{Fc}
\end{figure}

The shape of the potential \rf{sing1} for $m = 1$, $n=2$ is shown in Fig. \ref{Fchaotic} for $\delta = 10^{{-3}}$ and $\vp < 20$.
The potential  \rf{sing1} for $m = 1$, $n=2$ after stabilization of the axion field $\theta = 0$ \rf{sing2}  is shown in Fig. \ref{Fchaotic} for $\delta = 10^{{-3}}$ and $\vp < 10$.
\begin{table}[H] 
\begin{center}
\begin{tabular}{ |p{1cm}||p{1.5cm}|p{1.5cm}|p{1.5cm}|p{1.5cm}|  }
 \hline
  $\delta$ &  $ 10^{-4}$ & $5\times 10^{-4}$ & $ 10^{-3}$ & $2\times 10^{-3}$  \\
 \hline
 $n_{s}$ & 0.9656 & 0.9747 &  0.9847 & 1.002 \\
  \hline
 $r$ & 0.0015 &  0.0023 & 0.0039 & 0.0098\\
 \hline
 \end{tabular}
\end{center}
\caption{\footnotesize Values of $n_{s}$ and $r$ for  for the model \rf{sing2} with $m=1, n=2$, $\alpha = 1/3$, $N = 55$,  and various $\delta$. }
\label{k2}
\end{table}
\noindent These results show that even a relatively weak logarithmic singularity at the boundary of the moduli space \rf{sing1} is sufficient to increase $n_{s}$ from $\sim 0.967$ all the way to $n_{s} = 1$ while keeping $r$ well within the present observational constraints.

\subsubsection{Singular E-models}\label{sse}
The simplest E-model with a logarithmic singularity has a potential
\be
V(T, \bar T) = V_0\  \, \left (1  - {T+\bar T\over 2} + \delta  \,  \ln\Big ({2\over  T+ \bar T}\Big) \right)^{2} \ .
\label{Esing1}\ee
\begin{figure}[H]
\begin{center}
\includegraphics[scale=0.35]{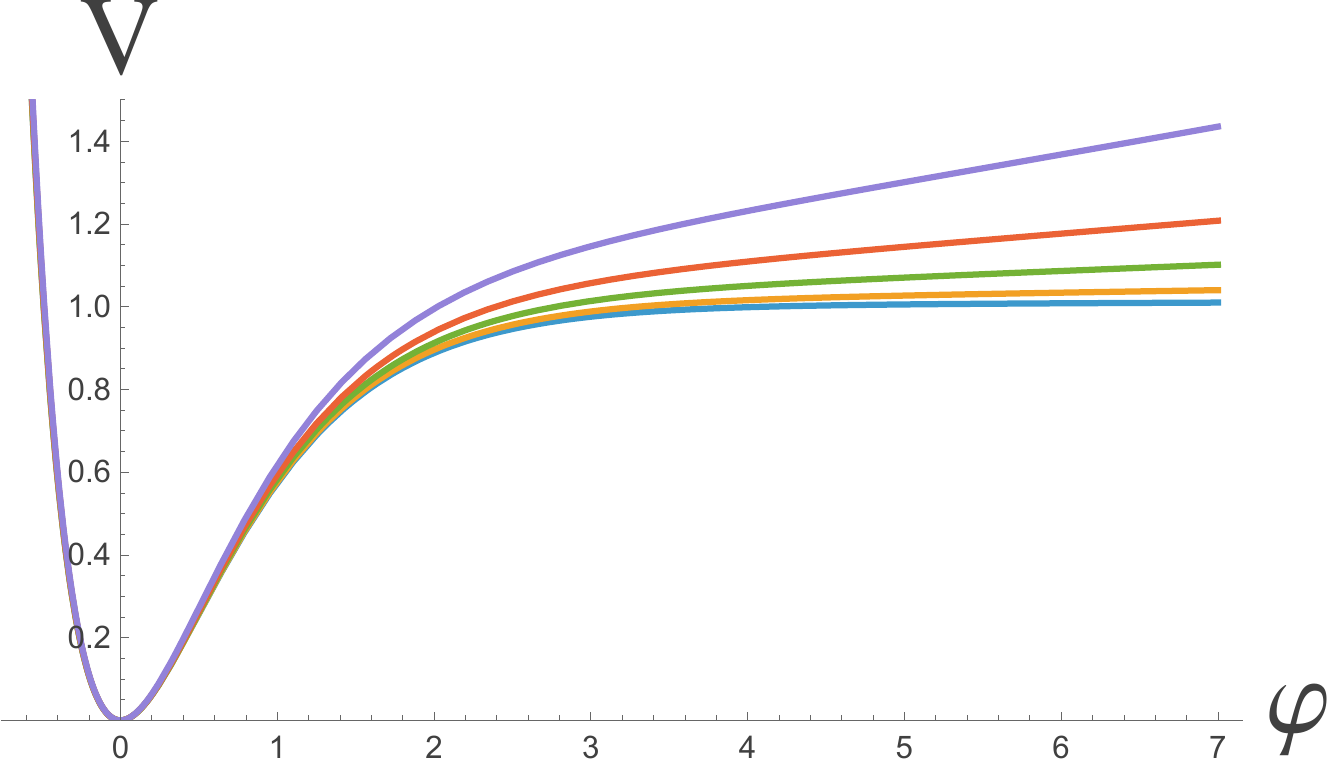}
\end{center}
%\vskip -0.5cm 
\caption{\footnotesize The potential \rf{Esing2} is shown in units of $V_{0}$  for $\alpha = 1/3$   and $\delta = 2\times 10^{-2}$ (upper curve), $\delta =10^{-2}$, $\delta=5\times 10^{-3}$,  $2\times 10^{-3}$, and $ 5\times 10^{-4}$ (lower curve). }
\label{EELL2}
\end{figure}
\noindent After stabilization of the axion field $T-\bar T = 0$, this potential  for $T=\bar T= e^{-{\sqrt 2\over {\sqrt {3\alpha}}} \vp}$ becomes
\be
V(\vp) = V_0\  \, \Big (1  - e^{-{\sqrt 2\over {\sqrt {3\alpha}}} \vp} + \delta  \, {\sqrt 2\over {\sqrt {3\alpha}}}\, \vp \Big )^{2} \ .
\label{Esing2}\ee
The plot of the potential is shown in Fig. \ref{EELL2}, and the values of $n_{s}$ and $R$ are given in Table \ref{EELT2}. 

\begin{table}[H] 
\begin{center}
\begin{tabular}{ |p{1cm}||p{1.5cm}|p{1.5cm}|p{1.5cm}|p{1.5cm}|p{1.5cm}|  }
 \hline
 $\delta$  &  $ 5\times 10^{-4}$ &  $ 2\times 10^{-3}$ & $5\times 10^{-3}$ & $ 10^{-2}$ & $2\times 10^{-2}$  \\
 \hline
 $n_{s}$ &  0.9657 & 0.9739 & 0.9794&  0.9879 & 0.9925 \\
  \hline
 $r$ &  $ 0.0015$ & 0.0017 & 0.0035 & 0.0073 & 0.0198\\
 \hline
 \end{tabular}
\end{center}
\caption{\footnotesize Values of $n_{s}$ and $r$ for   the model \rf{Esing2} with $\alpha = 1/3$,  $N = 55$ and various $\delta$. }
\label{EELT2}
\end{table}
\noindent One can also consider more general models, such as
\be
V(T, \bar T) = V_0\  \, \left (1  - {T+\bar T\over 2} + \delta  \,   \ln^{n}\Big ({2\over  T+ \bar T}\Big) \right )^{2}  \  .
\label{Esing1k}\ee
Just as in the simplest singular $T$-models \rf{sing2} studied in the previous section, the potential   \rf{Esing1k} at large $\vp$ is proportional to $\phi^{2n}$.

\subsection{Power-law singularity}

\subsubsection{Singular T-models}

The singular $\alpha$-attractor T-model introduced in  \cite{Linde:2017pwt,Linde:2018hmx} was already described at the beginning of this paper, see equation \rf{singL} and Fig. \ref{Wall}. We reproduce it here for convenience:
\be
V(Z, \bar Z) = V_0\ Z\bar Z \, \Big ( 1   + \, {\delta \over (1-Z\bar Z)} \Big ) \ .
\label{singLLL}\ee
\begin{figure}[H]
\begin{center}
\includegraphics[scale=0.42]{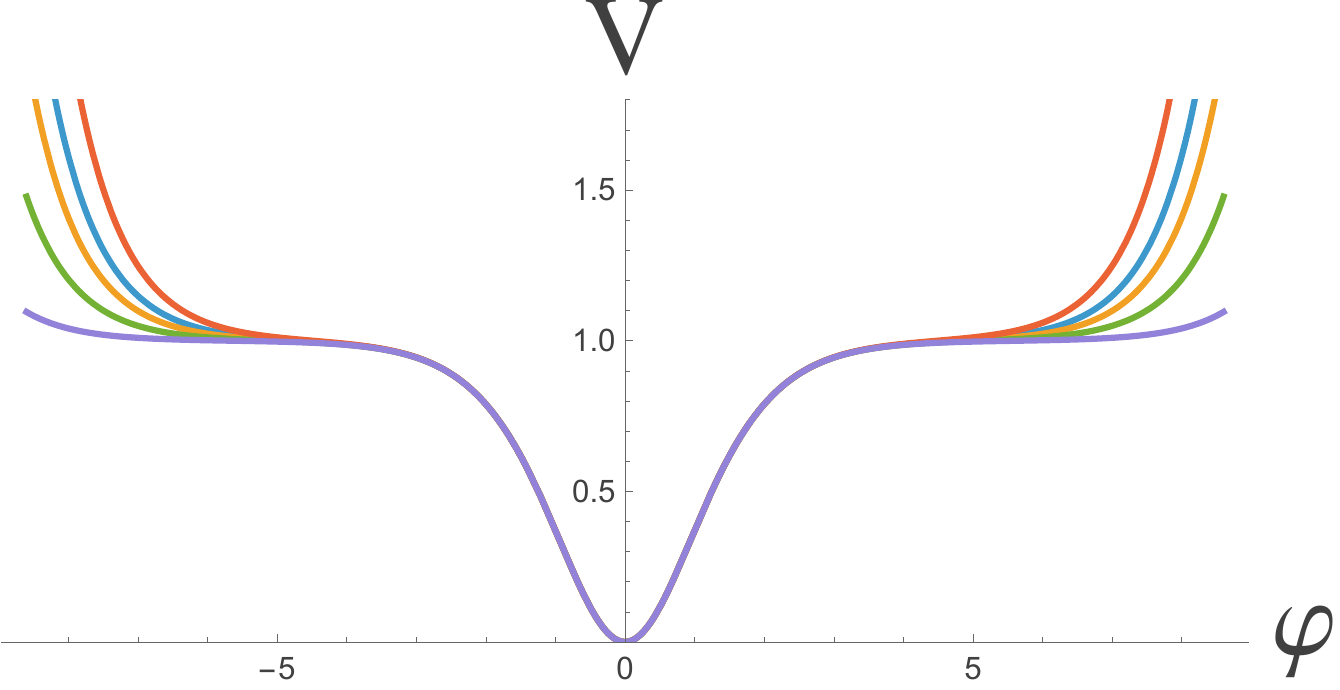}
\end{center}
%\vskip -0.5cm 
\caption{\footnotesize The potential \rf{VVV} shown in units of $V_{0}$, for $\alpha = 1/3$  and for $\delta = 5\times 10^{-5}$ (upper curve), $\delta = 3\times 10^{-5}$,  $\delta = 2\times 10^{-5}$, $\delta= 10^{-5}$,  and $\delta = 2\times 10^{-6}$ (lower curve). }
\label{F6}
\end{figure}
\noindent In canonical variables $\vp$, where $Z = e^{i \theta} \tanh {\vp\over \sqrt{6\alpha}}$, this potential, in the simplest case $m =1$, $n=1$ becomes
 \be
  V   = V_{0}  \left(\tanh^{2}{\varphi\over\sqrt {6 \alpha}} +\delta \sinh^{2}{\varphi\over\sqrt {6 \alpha}}\right)  \ .
\label{VVV}\ee
For  $\delta \ll 1$, this potential has a plateau, which ends at large $\vp$, where the potential becomes exponentially steep, see Fig. \ref{F6}. 
The results of the calculation of $n_{s}$ and $r$ for $\alpha = 1/3$ and $N = 55$ are presented in the Table \ref{k3}.
\begin{table}[H]
\begin{center}
\begin{tabular}{ |p{1cm}||p{1.5cm}|p{1.5cm}|p{1.5cm}|p{1.5cm}|p{1.5cm}|  }
 \hline
  $\delta$ & $2\times 10^{-6}$ &$ 10^{-5}$ & $2\times 10^{-5}$ & $3\times 10^{-5}$ & $5\times 10^{-5}$\\
 \hline
 $n_{s}$ & 0.9643 &0.9691 & 0.9774 & 0.9822 &0.9973\\
 \hline
 $r$ & 0.0014 &  0.0016 & 0.0018 & 0.0022 & 0.0032\\
 \hline
 \end{tabular}
\end{center}
\caption{\footnotesize Values of $n_{s}$ and $r$ for the model \rf{VVV} with $\alpha = 1/3$ and  $N = 55$.}
\label{k3}
\end{table}
\noindent Many of our numerical examples are given for $\alpha = 1/3$ because there are seven preferred values of $3\alpha$ in the context of extended supergravity models: $3\alpha = 1,2,3,...,7$.  The results for all of these values of $\alpha$ are qualitatively similar. For completeness, we present here the results for   $\alpha = 7/3$ and $N = 55$ in Table \ref{k4}:
\begin{table}[H]
\begin{center}
\begin{tabular}{ |p{1cm}||p{1.5cm}|p{1.5cm}|p{1.5cm}|p{1.5cm}|p{1.5cm}|  }
 \hline
  $\delta$ & $ 5\times 10^{-5}$ &$10^{-4}$ & $5\times 10^{-4}$ & $10^{-3}$ & $ 2\times 10^{-3}$\\
 \hline
 $n_{s}$ & 0.9640 &0.9646 & 0.0.9695 &0.9756 &0.9888\\
 \hline
 $r$ & 0.0091 &  0.0092& 0.0092& 0.0126 & 0.0180\\
 \hline
 \end{tabular}
\end{center}
\caption{\footnotesize Values of $n_{s}$ and $r$ for the model \rf{VVV} with $\alpha = 7/3$ and  $N = 55$.}
\label{k4}
\end{table}
\noindent Thus, we see that in the context of S-models one can easily match the latest ACT-SPT-DESI constraints  $n_{s}= 0.9728 \pm 0.0027$ and $r < 0.036$,

In this theory, for sufficiently small $\delta$, one has a long stage of inflation dominated by the simplest $\alpha$-attractor potential $ V_{0} \,  \tanh^{2\gamma}{\varphi\over\sqrt {6 \alpha}}$. However, in the large  $\varphi$ limit,  the potential \rf{VVV} is given by 
 \be 
  V  =  {V_{0}\, \delta\,  \over 4}\  e^{\sqrt {2\over 3 \alpha}\,\vp}.
\label{VVV2a}\ee
One may also consider  generalized versions of the theory   \rf{singLLL}, such as
\be
V(Z, \bar Z) = V_0\ (Z\bar Z)^{\gamma} \, \Big ( 1   + \, { \delta  \over (1-Z\bar Z)^{\beta}} \Big ) = V_{0}  \tanh^{2\gamma }{\varphi\over\sqrt {6 \alpha}}\left(1 +\delta \cosh^{2\beta}{\varphi\over\sqrt {6 \alpha}}\right) \ .
\label{singT}\ee
Here $\beta, \gamma$ are some positive (not necessarily integer) numbers.  In the large $\vp$ limit one has 
\be 
  V  =  {V_{0}\, \delta\,  \over 2^{2\beta}}\  e^{\beta\sqrt {2\over 3 \alpha}\,\vp}.
\label{VVV2b}\ee

\subsubsection{Singular E-models}

\begin{figure}[H]
\begin{center}
\includegraphics[scale=0.42]{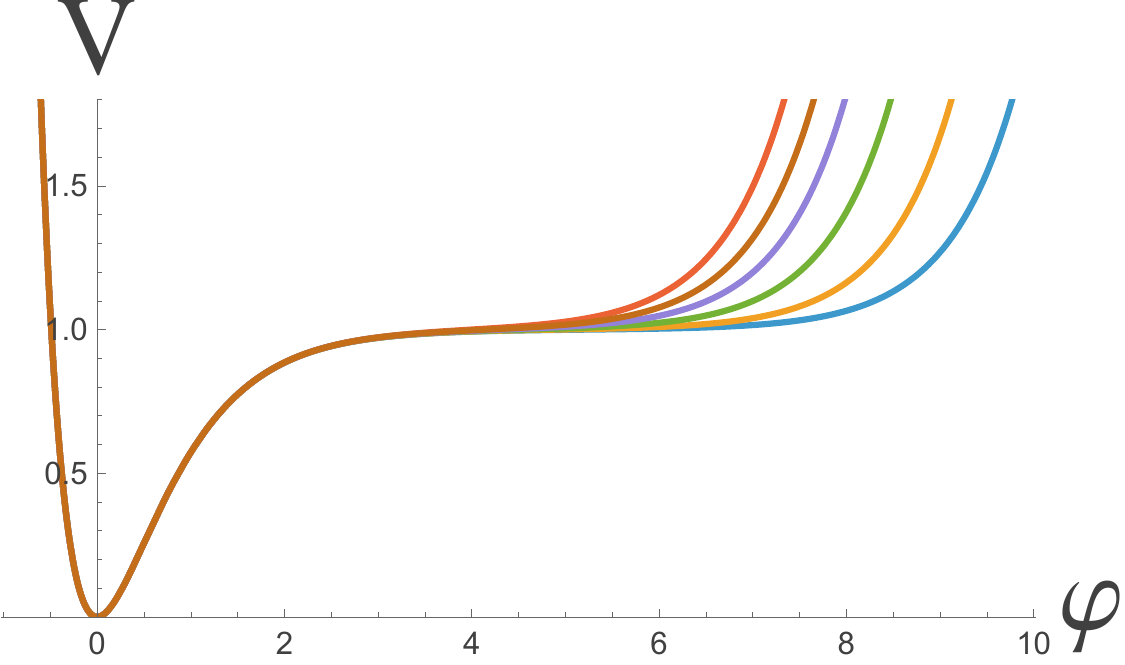}
\end{center}
%\vskip -0.5cm 
\caption{\footnotesize The potential \rf{EsingT2} is shown in units of $V_{0}$  for $\alpha = 1/3$,  and $\delta =   10^{-5}$ (upper curve), $\delta  =8\times 10^{-6}$, $\delta =5\times 10^{-6}$,  $\delta =2\times 10^{-6}$, and $\delta = 5\times 10^{-7}$ (lower curve). }
\label{EELF}
\end{figure}

The simplest E-model with a power-law singularity has a potential
\be
V(T, \bar T) = V_0\  \, \Bigl(1- {T+\bar T\over 2}\Bigr)^{2}  \Big (1+ \delta  \    \Big ({T+ \bar T\over 2}\Big)^{{-1}} \Big ) \ ,
\label{EsingT}\ee
where $ |1  -  T|^{2} \equiv (1  -  T)(1-\bar T)$. For $T=\bar T= e^{-{\sqrt 2\over {\sqrt {3\alpha}}} \vp}$ this potential becomes
\be
V(\vp) = V_0\  \, \Big(1  - e^{-{\sqrt 2\over {\sqrt {3\alpha}}} \vp} \Big)^{2}\, \Bigl(1+ \delta   \   e^{{\sqrt 2\over {\sqrt {3\alpha}}} \vp}\Bigr)  \ .
\label{EsingT2}\ee
Asymptotic behavior of $V(\vp)$ in this model in the large $\vp$ limit is the same as in \rf{sing4}.
 The plot of the potential is shown in Fig. \ref{EELF}, and the values of $n_{s}$ and $R$ are given in Table \ref{EELT}. 
\begin{table}[H] 
\begin{center}
\begin{tabular}{ |p{1cm}||p{1.6cm}|p{1.6cm}|p{1.6cm}|p{1.6cm}|p{1.6cm}| p{1.6cm}|  }
 \hline
  $\delta$ & $8\times 10^{-7}$ &$2\times 10^{-6}$ & $5\times 10^{-6}$ & $ 10^{-5}$ & $ 1.6 \times 10^{-5}$& $ 2.5\times 10^{-5}$\\
 \hline
 $n_{s}$ & 0.9647 &0.9661& 0.9698 &0.9763 &0.9847&0.9991\\
 \hline
 $r$ & 0.0013&  0.0014& 0.0015& 0.0018 & 0.0023&0.0032\\
 \hline
 \end{tabular}
\end{center}
\caption{\footnotesize Values of $n_{s}$ and $r$ for  the model \rf{EsingT2} with  $\alpha = 1/3$, $N = 55$,  and various $\delta $. }
\label{EELT}
\end{table}
\noindent One may consider various generalizations of this model, such as
\be
V(T, \bar T) = V_0\  \, \Bigl(1- {T+\bar T\over 2}\Bigr)^{2n} \  \Big (1+ \delta  \    \Big ({T+ \bar T\over 2}\Big)^{{-\beta}} \Big )  \ ,
\label{EsingT3}\ee
where  $\beta$ can be any positive number.  This yields
\be
V(\vp) = V_0\  \, \Big(1  - e^{-{\sqrt 2\over {\sqrt {3\alpha}}} \vp} \Big)^{2n}\, \Bigl(1+ \delta   \   e^{\beta {\sqrt 2\over {\sqrt {3\alpha}}} \vp}\Bigr)  \ .
\label{EsingT4}\ee
In the large  $\varphi$ limit,  the potential \rf{EsingT4} is given by 
 \be 
  V  =  {V_{0}\, \delta\,  \over 4}\  e^{\beta \sqrt {2\over 3 \alpha}\,\vp}.
\label{VVV2}\ee
This result coincides with the corresponding result for T-models \rf{VVV2}. 

\subsection{S-models with $SL(2,\mathbb{Z})$ symmetry}

In $SL(2,\mathbb{Z})$ models where $T=-i\tau$ \cite{Kallosh:2024ymt}  inflation takes place  at $ {\rm Im} \tau \to 0 $ where the $SL(2,\mathbb{Z})$ invariant  $j$-function at large values of the canonical field $\vp$ and at stabilized axions has the following asymptotics
\be
|j(\tau)|^2   \to    e^{4\pi e^{\sqrt{2\over 3\alpha} \varphi} } \,.
\label{largej2}\ee
An example of the singular $SL(2,\mathbb{Z})$ model with a power-law  singularity at $ {\rm Im} \tau \to 0 $ can be given in the form
\be
V=
V_0\left(1-{\ln  j(i)^{2}
\over  \ln\Big[|j(\tau)|^2+A\, | j(\tau) -\overline {j(\tau)}|^{2} + j(i)^{2}\Big]}\right) \Bigl( 1+ \delta \ \ln^\beta \big(|j(\tau)|^2 + j(i)^{2}\big)\Bigr) \ .
\label{beta2}
\ee
At $\delta=0$ this equation represents a non-singular $SL(2,\mathbb{Z})$ invariant $\alpha$-attractor potential with axions stabilized by the term $A\, | j(\tau) -\overline {j(\tau)}|^{2}$, see eq. (2.4) in \cite{Carrasco:2025rud}.  The singular term $\delta  \ \ln^\beta \Big[|j(\tau)|^2 + j(i)^{2}\Big]$ grows as $ \delta \, (4\pi )^{\beta }\, e^{\beta \sqrt{2\over 3\alpha} \varphi}$ at large $\vp$. The full potential in the large $\vp$ limit is
\be
V(\vp) = V_{0}\Bigl(1-{\ln  j(i)\over 2\pi} e^{-\sqrt{2\over 3\alpha}\vp }\Bigr) \Bigl(1+  \delta \, (4\pi )^{\beta}\, e^{\beta \sqrt{2\over 3\alpha} \varphi}\Bigr) \ .
\label{beta3}\ee
In the absence of the last term, this theory has the standard $\alpha$-attractor predictions since the numerical factor ${\ln  j(i)\over 2\pi}$ can be absorbed in the redefinition (shift) of the field $\vp$. The last term allows for a significant increase in $n_{s}$ controlled by the parameter $\delta$, just as in all other previously discussed models.

\subsection{  S-models of $\a$-attractors related to Fibre inflation }\label{fibre}
 
 \

 \noindent Interesting examples of quantum corrections lifting the plateau potentials are known in the context of Fibre inflation in string theory \cite{Cicoli:2008gp,Cicoli:2024bxw}.  It was also shown in \cite{Kallosh:2017wku} that Fibre inflation models 
 can be effectively described as a particular case of  supergravity $\a$-attractors in cases of $\a=1/2$ or $\a=2$.

In the case associated with $\a=1/2$ attractors in \cite{Cicoli:2008gp}, the  single field effective Fibre model potential is given as 
\be
V(\vp)\approx V_0 \left  (3  - 4 e^{-{\vp\over \sqrt 3} } + e^{-4{\vp\over \sqrt 3} } +R \  e^{{2\vp\over \sqrt 3} }\right  ) \ ,
\label{Fer1}\ee
where $R\ll1$.  
This model, when viewed as a $\a=1/2$ attractor, has a kinetic term given in eq. \rf{kin} as 
\be
{3\alpha\over 4} \, {\partial T \partial \bar T\over ({\rm Re} \,  T )^2}\Big |_{\a=1/2} = {3\over 8} \, {\partial T \partial \bar T\over ({\rm Re} \,  T )^{2}} \ .
\ee
We will now show that this potential as a function of geometric moduli $(T, \bar T)$ has the form of a singular $\a=1/2$ attractor
\be
V(T, \bar T)= V_0 \left  (3 -4 \left({T+\bar T\over 2}\right)^{1/2} + \left({T+\bar T\over 2}\right)^2 +R \left({T+\bar T\over 2}\right)^{-1}\right  ) \ .
\label{Fer}\ee
This potential is singular at the boundary when $R\neq 0$.
One can check that at $T=\bar T= e^{-{2\over {\sqrt 3}} \vp}$ the potential  \rf{Fer} coincides with the expression in equation \rf{Fer1}.
The values of $n_s, r$ were computed in  \cite{Cicoli:2008gp}  for some choices of parameters. The benchmark models for  $R\approx 2\times 10^{-6}$ and $N=57$ gave $n_s\approx 0.97, r\approx 6\times 10^{-3}$ and for $N=53$ $n_s\approx 0.967, r\approx 6 \times 10^{-3}$.

In the case associated with $\a=2$ attractors in \cite{Cicoli:2024bxw}, a single field effective Fibre model potential in eq. (2.25) in \cite{Cicoli:2024bxw} is
\be
V(\vp)= V_0 \left  (E +A e^{-{4\vp\over \sqrt 3} } -B e^{-{\vp\over \sqrt 3} } +C e^{{2\vp\over \sqrt 3} }\right  ) \ .
\label{Cic1}\ee
This model, when viewed as a $\a=2$ attractor, has a  kinetic term given in eq. \rf{kin} as 
\be
{3\alpha\over 4} \, {\partial T \partial \bar T\over ({\rm Re} \,  T )^2}\Big |_{\a=2} = {3\over 2} \, {\partial T \partial \bar T\over ({\rm Re} \,  T )^{2}} \ .
\ee
We find that this potential as a function of geometric moduli $(T, \bar T)$ has the form of a singular $\a=2$ attractor
\be
V(T, \bar T)= V_0 \left  (E + A \left({T+\bar T\over 2}\right)^4 - B \left({T+\bar T\over 2}\right) +C \left({T+\bar T\over 2}\right)^{-2}\right  ) \ .
\label{Cic}\ee
The slice of it at $T=\bar T= e^{-{\sqrt{2\over 3\a}} \vp}$ and at $\a=2$ leads to a single field effective Fibre model potential in eq. (2.25) in \cite{Cicoli:2024bxw}, which is given here  in eq. \rf{Cic1}.

When $C\neq 0$ $\a$-attractor is singular. The constant $C$ describing the term singular at the boundary in the potential \rf{Cic} is very small in examples studied in \cite{Cicoli:2024bxw}. It bends the plateau upwards when $C\sim 10^{-5}$   and it  removes the plateauat $C\sim 10^{-3}$.
The benchmark models in \cite{Cicoli:2024bxw} have $n_s$ close to the Planck-preferred values.

More recently, after ACT, SPT, DESI results, related models were studied in \cite{Leontaris:2025hly}, where the effective single-field Fibre inflation   potential is
\be
V(\vp)= C_0 \left  (R_{LVS} +  R_0  e^{-{2\vp\over \sqrt 3} }-  e^{-{\vp\over \sqrt 3} } +R_1 e^{{\vp\over \sqrt 3} } +R_2(\vp) e^{{2\vp\over \sqrt 3} }\right  )
\label{Leon1}\ee
Here $R_2(\vp) $ exponentially fast approaches a constant at large $\vp$.

This model can  also be interpreted as a singular $\a=2$ attractor: we find the relevant $V(T, \bar T)$ potential, which is singular at the boundary
\be
V(T, \bar T)= C_0 \left  (R_{LVS} +  R_0  \left({T+\bar T\over 2}\right)^2 -  \left({T+\bar T\over 2}\right) +R_1 \left({T+\bar T\over 2}\right)^{-1} +R_2(T, \bar T) \left({T+\bar T\over 2}\right)^{-2}\right  )
\label{Leon}\ee
At $T=\bar T= e^{-{\sqrt{2\over 3\a}} \vp}$ and   $\a=2$ the $V(T, \bar T)$ potential in eq. \rf{Leon} coincides with the single field effective Fibre inflation potential given in eq. 
(3.32)  of    \cite{Leontaris:2025hly}, or here in eq. \rf{Leon1}. 

For a specific choice of parameters of this model, which includes very small values of singular terms near the boundary
\be
R_1 \approx 2\times 10^{-4}\, , \qquad R_2\approx - 2\times 10^{-5} \ ,
\ee
the authors of \cite{Leontaris:2025hly} have found $n_s$ and $r$ predictions compatible with ACT, SPT, DESI. Their 3 benchmark models  include $n_s\approx 0.974,\, 0.975,\, 0.975$ with $r\approx 5.4 \times 10^{-3}, \, 4 \times 10^{-3}, \, 4\times  10^{-3}$.

 \subsection{    S-models   related to  deformed $\a$-attractor models} \label{defa}
 Recently, a set of cosmological deformed $\a$-attractors models was proposed in \cite{Ellis:2025zrf}. These models were defined via a potential of the canonical field $\vp$. One of the models was a deformed E-model  model with a potential
 \be
 V(\vp)= V_0 \left (\kappa \Big(1-\cosh \Big ( \sqrt {2\over 3\a} \vp\Big) \Big) + \sinh \Big ( \sqrt {2\over 3\a} \vp \Big)  \right)^k \ ,
 \label{def1}\ee
 where at $\kappa=1$ the  E-model $\a$-attractor   is restored and the potential becomes 
 \be
 V(\vp)= V_0 \Big ( 1-e^{ -\sqrt {2\over 3\a} \vp } \Big)^k \ .
 \label{def1noR}\ee
 The second was a deformed T-model with the potential
 \be
 V(\vp)=  {V_0\over 4} \Big( \Big [1+\kappa -(\kappa-1)  \cosh \Big ( \sqrt {2\over 3\a} \vp \Big)\Big]^{2}  \times  \tanh^k \Big ( {\vp \over \sqrt { 6 \a} } \Big)  \Big) \ .
 \label{def2} \ee 
 At $\kappa=1$ this is an inflationary potential of the T-model $\a$-attractor 
\be
 V(\vp)=  V_0  \tanh^k \Big ( {\vp \over \sqrt { 6\a} } \Big)  \ .
 \label{def2noR} \ee 
The supersymmetric embedding of these models in  \cite{Ellis:2025zrf} was given  in  supergravity  with two superfields, $T$ and $\Phi$,  where the  generalized $\a$-attractor \K potential  is taken in the form
\be
K= - 3\, \a\, \ln \Big (T+\bar T - {\Phi \bar \Phi\over 3}\Big  ) \ .
\label{noscale}\ee
The superpotentials are presented in a form $W(T, \Phi)$ and carry the information on $k$ and $\kappa$. The potentials were computed for various choices of  $W(T, \Phi)$. The final potentials $V(T, \bar T)$ and $V(\Phi, \bar \Phi) $ are taken at fixed $\Phi$ or at fixed $T$. These potentials along the inflationary trajectory lead to equations \rf{def1} and \rf{def2}. The superpotentials $W(T, \Phi)$ and potentials $V(T, \bar T)$ and $V(\Phi, \bar \Phi) $  are rather complicated and do not have any obvious relation to geometric  S-models discussed in our paper.

However, as we will show now, it is possible to embed the models \rf{def1}, \rf{def2} into the class of singular $\a$-attractor models with \K potentials  \rf{K}. As in the case with Fibre inflation models, we need to supply the potentials depending on $Z, \bar Z$ or $T, \bar T$, which coincide with \rf{def1}, \rf{def2} after the axion stabilization. Thus, we take  a potential that is singular at the boundary $T+\bar T\to 0$ and the \K potential
\be
 K= - 3\a \ln  (T+\bar T)\, ,  \quad V(T, \bar T)= V_0 \left (1-\delta -\Big({T+\bar T\over 2}\Big) + {\delta\over 2}\Big[ \Big ({T+\bar T\over 2} \Big )  + \Big ({T+\bar T\over 2} \Big )^{-1} \Big]\right)^k
 \label{def1E}\ee
 This is the $\a$-attractor E-model containing a singular term   $ \delta \Big ({T+\bar T\over 2} \Big )^{-1}$. One can show that after axion stabilization with $T=\bar T = e^{-{\sqrt{2\over 3\a}} \vp}$,  this potential coincides with  \rf{def1} for $\kappa=1-\delta$. 
 In the large $\vp$ limit, the potential grows as   $V_{0} \, ({\delta\over 2})^{k} \, e^{k\sqrt {2\over 3\a} \vp }$.
 
The S-model representation of the potential \rf{def2} is
\be
 K(Z, \bar Z)=-3\a \log(1-Z\bar Z)\, ,  \qquad V(Z, \bar Z)= V_0  \ (Z\bar Z)^{k\over 2}  \left (1 + \delta \, {Z\bar Z \over 1-Z\bar Z }\right)^2 \ .
 \label{def1T}\ee
 This is the usual $\a$-attractor T-model deformed by the singular term $\delta \, {Z\bar Z \over 1-Z\bar Z }$.
One can show that this potential with the stabilized axion and   $Z = \bar Z= \tanh \Big ( {\vp \over \sqrt { 6 \a} }\Big) $ is equal to the potential in eq. \rf{def2}. In the large $\vp$ limit this potential grows as $V_{0} {\delta^{2}\over 4} e^{2\sqrt {2\over 3\a} \vp }$.

Thus, we have constructed new S-models of singular $\a$-attractors associated with  Fibre inflation in \cite{Leontaris:2025hly} and deformed E- and T-models  in \cite{Ellis:2025zrf}. The cosmological predictions of thess models are similar to the predictions of our S-models. In all cases, the models with the potential singular near the boundary allow significantly higher values of $n_s$. This is good news with regard to ACT, SPT, DESI data.

\section{S-models and the problem of initial conditions for inflation}\label{in}

In the old Big Bang theory, cosmological evolution was supposed to be approximately adiabatic, and the total number of particles in the universe was supposed to be approximately conserved. From this assumption, one could show that at the Planckian time $t = O(1)$  when the density of the hot universe was Planckian $\rho = O(1)$, it consisted of about $10^{{90}}$ causally disconnected parts of a Planckian size $l = O(1)$, each of which contained only $O(1)$ elementary particles. In this case, it was hard to explain large-scale flatness, homogeneity and isotropy of the universe, and the origin of the exponentially large number of particles in the universe.

To explain how inflationary theory solves this problem, let us consider a closed universe of the smallest initial size $l = O(1)$, which emerges in a state
with the Planck density $\rho = O(1)$. This condition implies that the sum of the kinetic energy density, gradient energy density, and the potential energy density is of the order unity, in Planck units:\, ${1\over 2} \dot\phi^2 + {1\over 2} (\partial_i\phi)^2 +V(\phi) \sim 1$.

There are no {\it a priori} constraints on
the initial value of the scalar field in this domain, except for the
constraint ${1\over 2} \dot\phi^2 + {1\over 2} (\partial_i\phi)^2 +V(\phi) \sim
1$.  Consider, for a moment, a theory with $V(\phi) = const$. This theory
is invariant under the {\it shift symmetry}  $\phi\to \phi + c$. Therefore, in such a
theory {\it all} initial values of the homogeneous component of the scalar field
$\phi$ are equally probable.  

The only constraint on the initial value of the field appears if the effective potential is not constant but grows and becomes greater
than the Planck density at $\phi > \phi_p$, where  $V(\phi_p) = 1$. This
constraint implies that $\phi \lesssim \phi_p$,
but there is no model-independent reason to expect that initially
$\phi$ must be much smaller than $\phi_p$. In the context of the realistic version of the model with the potential $ {m^{2}\over 2} \phi^{2} $  with $m = O(10^{{-5}})$ this implies that the upper constraint on the initial value of the field $\phi$ is $\phi \lesssim 1/m \sim 10^{5}$.

Thus, we expect that typical initial conditions correspond to
${1\over 2}
\dot\phi^2 \sim {1\over 2} (\partial_i\phi)^2\sim V(\phi) = O(1)$.
If ${1\over 2} \dot\phi^2 + {1\over 2} (\partial_i\phi)^2
 \lesssim V(\phi)$ in the part of the universe under consideration, one can show that inflation begins,
and then within the Planck time the terms  ${1\over 2} \dot\phi^2$ and ${1\over 2}
(\partial_i\phi)^2$ become much smaller than $V(\phi)$, which ensures
continuation of inflation.  The conclusion is that inflation
occurs under rather natural initial conditions if it can begin at $V(\phi)
= O(1)$ \cite{Linde:1983gd,Linde:1985ub,Linde:2005ht}.  

This conclusion matches the results of the theory of quantum creation of a closed universe ``from nothing''. According to \cite{Linde:1983mx,Linde:1984ir,Vilenkin:1984wp}, the probability of quantum creation of a closed inflationary universe with potential $V(\phi)$ is given by
\be\label{prob}
P \sim  e^{-{24\pi^{2}\over V}} \ .
\ee
This implies that quantum creation of the universe with sufficiently large $V$ is not suppressed. This condition can be easily met by the simplest chaotic inflation models with $V \sim \phi^{n}$ \cite{Linde:1983gd}.

On the other hand, quantum creation of the universe with a plateau potential of height  $V_{0}\sim 10^{{-10}}$ (which is the case for the simplest single-field $\alpha$-attractor models), the probability becomes exponentially suppressed by a factor $P \sim 10^{-10^{12}}$. 

To solve this problem, one may add to the $\alpha$-attractor theory a second field $\chi$ with a potential ${m^{2}\over 2} \chi^{2}$. Then inflation driven by the field $\chi$ may begin at  ${m^{2}\over 2} \chi^{2} = O(1)$, as in the simplest chaotic inflation model discussed above. If the field $\vp$ initially was at the plateau of its potential $V(\vp)$, it does not move until the end of the first stage of inflation driven by the field $\chi$. After that, the stage of inflation driven by the field $\vp$ begins, and if this stage is long enough, it determines the formation of the large-scale structure and CMB perturbations in the observable part of the universe. Various versions of this scenario have been discussed in \cite{Linde:1987yb,Linde:2014nna,Carrasco:2015rva,Dimopoulos:2016yep,Linde:2017pwt}, and its validity has been confirmed by detailed numerical calculations in \cite{Corman:2022alv}.

There are many other, more sophisticated ways to solve this problem \cite{Carrasco:2015rva,East:2015ggf,Kleban:2016sqm,Clough:2016ymm,Linde:2017pwt,Clough:2017efm,Aurrekoetxea:2019fhr,Creminelli:2020zvc,Joana:2020rxm,Joana:2022pzo,Corman:2022alv,Elley:2024alx,Aurrekoetxea:2024mdy,Joana:2024ltg}.  In particular, it was noticed  in 
\cite{Linde:2017pwt,Linde:2018hmx} that the solution of this problem becomes nearly trivial in the $\a$-attractor S-models.

Let us consider first the S-models  with a logarithmic singularity \rf{sing1} discussed in section \ref{ls}.  At small and intermediate values of the inflaton field $\vp$ its potential is given by the standard T-model expression $=V_0\, \tanh^{2n} {\vp\over \sqrt{ 6 \a}} $. However, at large $\vp$ its potential is given by the standard monomial chaotic inflation potential $V  \sim  |\vp|^n$, see equation \rf{sing3}.  Similarly, the potential of the E-model \rf{Esing1k} at large positive $\vp$ grows as $\phi^{2n}$. 

Thus, inflation in both of these models can begin at very large $\phi$ just as in the simplest version of the chaotic inflation scenario. This solves the problem of initial conditions in these models.

The situation in the S-models with a power-law singularity is very similar, though slightly more nuanced. Following  \cite{Linde:2017pwt,Linde:2018hmx}, we will consider first the simplest T-model \rf{singLLL}. Just as before, at small and medium values of $\vp$, its potential  is $V=V_0\, \tanh^{2n} {\vp\over \sqrt{ 6 \a}} $, but at large $\vp$ it grows as $e^{\sqrt {2\over 3 \alpha}\,\vp}$, see \rf{VVV2a}.

Cosmological evolution in such potentials is well known  \cite{Liddle:1988tb}: 
\be\label{power}
a(t) = a_{0}\ t^{3\,\alpha} \ .
\ee
For  $\alpha > 1/3$, the power-law solution \rf{power} describes inflation with $\ddot a > 0$, i. e. inflation. In this class of models, just as in the models with the logarithmic singularity discussed in section \ref{ls}, inflation may already begin at the Planck density. This solves the problem of initial conditions in such models along the lines of \cite{Linde:1985ub}.  

For $\alpha = 1/3$ we have  $a \sim t$, $\ddot a = 0$. This regime shares many of the properties of inflation. In particular, the energy of the homogeneous component of the scalar field decreased as $a^{{-2}}$, i.e. much more slowly than the energy of dust $\sim a^{{-3}}$, of the relativistic gas $\sim a^{{-4}}$ and of the gradient energy of the scalar field.  This makes the solution of the problem of initial conditions addressed in \cite{Carrasco:2015rva,East:2015ggf,Kleban:2016sqm,Clough:2016ymm,Linde:2017pwt,Clough:2017efm,Aurrekoetxea:2019fhr,Creminelli:2020zvc,Joana:2020rxm,Joana:2022pzo,Corman:2022alv,Elley:2024alx,Aurrekoetxea:2024mdy,Joana:2024ltg} much simpler: If initially the energy of the homogeneous field was comparable to other types of energy, including the energy of inhomogeneities, then in an expanding universe the energy of the homogeneous field gradually starts to dominate. It continues to dominate until the field approaches the plateau of the potential, and the familiar $\alpha$-attractor inflationary regime begins. Thus, one may argue that the problem of initial conditions is solved not only for $\alpha > 1/3$, but for $\alpha = 1/3$ as well.

In generalized models of this type \rf{singT}, the potential grows at large $\vp$ as $e^{\beta\sqrt {2\over 3 \alpha}\,\vp}$, see \rf{VVV2b}.
In this class of models, the early stage of inflation is possible for $3\alpha > \beta^{2}$. Thus, in the models with small $\beta$, one can solve the problem of initial conditions even if $\alpha < 1/3$. The situation is very similar in the singular E-models \rf{EsingT} and \rf{EsingT3}, and in the $SL(2,\mathbb{Z})$-invariant model \rf{beta2}.

In all versions of Fibre inflation studied in section \ref{fibre}, the potential at large $\vp$ grows as $e^{2\phi\over \sqrt 3}$, which is also compatible with inflation at large $\vp$.

As for the deformed $\alpha$-attractors models discussed in section \ref{defa}, the problem of initial conditions is solved in the model \rf{def1} for $3\alpha \geq k^{2}$.  For the model \rf{def2}, the problem of initial condition is solved for $\alpha \geq 4/3$.

\section{Discussion}

In this paper, we discussed a generalized class of $\alpha$-attractors, S-models, which have a singularity at the boundary of the moduli space  \cite{Linde:2017pwt,Linde:2018hmx}.  This singular behavior may uplift the inflationary plateau at large values of the canonically normalized inflaton field. In particular, a logarithmic singularity leads to a power-law growth of the potential at large values of $\vp$, as in the simplest versions of the chaotic inflation scenario with $V \sim \vp^{n}$, see Figs. \ref{Fa}, \ref{Fchaotic}.  Meanwhile, power-law (pole) singularities give rise to potentials that grow exponentially at large $\vp$, see Figs. \ref{Wall}, \ref{finitePla}. In many cases discussed in this paper, such potentials support inflation at arbitrarily large values of the potential, which helps to solve the problem of initial conditions for inflation along the lines of \cite{Linde:1983gd,Linde:1985ub,Linde:2005ht}. We found that these conclusions remain valid for several other closely related inflationary models, such as Fibre inflation \cite{Cicoli:2008gp,Cicoli:2024bxw,Leontaris:2025hly} and deformed $\alpha$-attractors  \cite{Ellis:2025zrf}.

The strength of the modification of the $\alpha$-attractor potential related to the singular terms in our models is controlled by a small parameter $\delta$. In many of the models considered in this paper, it is sufficient to take $\delta \sim 10^{{-5}}$ to move $n_{s}$ towards the area favored by the recent constraints based on combining CMB and DESI results.

The resulting situation is somewhat similar to what we have encountered with the invention of $\alpha$-attractors. Prior to that, there was a set of models such as the Starobinsky model, the Higgs inflation model, $\xi$-attractors and conformal attractors, which made identical predictions for $n_{s}$ and $r$. Meanwhile, $\alpha$-attractors made the same prediction for $n_{s}$,  but allowed to account for all possible values of the tensor-to-scalar ratio $r= 12\alpha/N^{2}$.

Now we are in a similar situation where we may need flexibility to describe a new, extended range of $n_{s}$. This flexibility can be provided by P-models  \cite{Kallosh:2022feu}, but an even greater degree of freedom is provided by S-models: By a small change of the parameter $\delta$, one can cover a broad range of $n_{s}$ from the familiar $\alpha$-attractor prediction $n_{s} = 1-2/N$ to the Harrison-Zeldovich value $n_{s} = 1$. Thus, now we have a new way to address the problem of initial conditions for inflation and to match observational data within this theoretical framework.

 \section*{Acknowledgement}
We are grateful to  A. Achucarro, L.~Balkenhol, S.~Ferrara, E.~Ferreira, L.~Knox, E.~McDonough, M.~Olechowski, S.~Pokorski, D.~Roest, 
T.~Terada, T.~Wrase, and Y.~Yamada for many stimulating discussions.  This work is supported by Leinweber Institute for Theoretical Physics at Stanford and by NSF Grant PHY-2310429.

\appendix

\section{Streamlined supergravity version of singular $\a$-attractors}

The textbook N=1 supergravity has an F-term  potential depending on a superpotential $W(z_i)$ and a \K potential $K(z^i, \bar z^{\bar i})$, with the scalar potential  $V(z^i, \bar z^{\bar i})=e^K (|DW|^2 - 3 |W|^2)$. In this approach, it is not always easy to find the potential $V(z^i, \bar z^{\bar i})$ with the required properties. We show that in supergravity with a nilpotent superfield and {\it with any  \K potential}  $K(z^i, \bar z^{\bar i} )$ one can obtain {\it any desired potential}  $V(z^i, \bar z^{\bar i})$ by a proper choice of the \K metric of the nilpotent superfield.

The streamlined supergravity \cite{Kallosh:2025dac} depends on $n$ physical chiral superfields and one nilpotent superfield. For the models with  one physical scalar  $(z, \bar z)$  and a scalar $s$ from the nilpotent superfield $(s, \bar s)$, streamlined supergravity has the following  \K potential and superpotential:
\be
K(z, \bar z,  s, \bar s)=K(z, \bar z ) + {F_s \bar F_{\bar s}\over e^{-K(z, \bar z) } V(z, \bar z)+ |W_0|^2 (3-K^{z \bar k} K_zK_{\bar z}) } \  
   s \,\bar s\, , \quad 
W= W_0 + F_s \, s  \ .
\label{KW}\ee
The resulting bosonic action of  scalars is
\be
{ {\cal L} (z, \bar z)\over \sqrt{-g}} =  {R\over 2} + K_{z \bar z} (z,  \bar z)\, \partial  z \partial \bar z + K_{s\bar s}(z, \bar z)\, \partial  s \, \partial \bar {s}  - V(z, \bar z)  \ ,
\label{Baction}\ee
where the \K potential $K(z, \bar z)$ and potential $V(z, \bar z)$ are arbitrary. 
A complete supersymmetric version of these models with fermions is presented in a unitary gauge in \cite{Kallosh:2015sea} and before gauge fixing local non-linearly realized supersymmetry in \cite{Kallosh:2015tea}.
Various versions of this construction in the context of $\alpha$-attractors have been previously developed in \cite{Achucarro:2017ing,Yamada:2018nsk,Linde:2018hmx,Kallosh:2022vha,Kallosh:2025jsb}.  In particluar, 
the streamlined supergravity version of  $\alpha$-attractors in disk variables corresponds to  
\begin{align}
K(Z, \bar Z, s, \bar s)=&-3\alpha\ln(1-Z\bar Z)+\frac{F_s \bar F_{\bar s}}{(1-Z\bar Z)^{3\alpha} \, V(Z, \bar Z) +3|W_0|^2(1-\alpha Z\bar Z)} s\bar s  \ .
\label{G}\end{align}
The bosonic action following from this supersymmetric construction is 
\be
{ {\cal L} (Z, \bar Z)\over \sqrt{-g}} =  {R\over 2} - {3\alpha} \, {\partial Z \partial \bar Z\over (1-Z\bar Z)^2}-  V(Z, \bar Z)   \ .
\label{hyper2Z}\ee 
Here $V(Z, \bar Z)$ can be any potential, singular or non-singular, with or without the axion stabilization. 
To stabilize the axion without affecting the inflaton potential, one can add to the potential $V(Z,\bar Z)$ a term $A(Z,\bar Z) (Z-\bar Z)^{2n}$, which vanishes along the inflaton direction $Z = \bar Z$. The function  $A(Z,\bar Z) $ can be chosen to achieve strong axion stabilization for all values of the inflaton field $\vp$.

In half-plane variables $T$, one has
\be
K(T, \bar T,  s, \bar s)=-3\a \ln(T+\bar T)+ \frac{F_s \bar F_{\bar s}}{(T+\bar T)^{3\alpha} V(T,\bar T) +3|W_0|^2(1-\alpha)}  \, s \,\bar s \ .
\label{Kexample}\ee
In this case, the bosonic action is 
\be
{ {\cal L} (T, \bar T)\over \sqrt{-g}} =  {R\over 2} - {3\alpha\over 4} \, {\partial T \partial \bar T\over ({\rm Re} \,  T )^2}-  V(T, \bar T)  \ .
\label{hyper2}\ee
For axion stabilization, one can add to $V(T, \bar T)$ a stabilizing function $B(T, \bar T) (T-\bar T)^{2n}$, which does not affect the potential in the inflaton direction $T = \bar T$.

\bibliographystyle{JHEP}
\bibliography{lindekalloshrefs}
\end{document}